\documentclass[12pt]{article}
\usepackage{amsmath}
\usepackage{amssymb}
\usepackage{amsfonts}
\usepackage{epsfig}

\begin{document}

\title{Searching for the deformation-stability fundamental length (or
fundamental time)}
\date{}
\author{R. Vilela Mendes\thanks{%
CMAF, Instituto para a Investiga\c{c}\~{a}o Interdisciplinar, Av. Gama Pinto
2, 1649-003 Lisboa, Portugal, vilela@cii.fc.ul.pt,
http://label2.ist.utl.pt/vilela/} \thanks{%
IPFN - EURATOM/IST Association, Instituto Superior T\'{e}cnico, Av. Rovisco
Pais 1, 1049-001 Lisboa, Portugal}}
\maketitle
\tableofcontents

\begin{abstract}
The existence of a fundamental length (or fundamental time) has been
conjectured in many contexts. However, the "stability of physical theories
principle" seems to be the one that provides, through the tools of algebraic
deformation theory, an unambiguous derivation of \ the stable structures
that Nature might have chosen for its algebraic framework. It is well-known
that $c$ and $\hbar $ are the deformation parameters that stabilize the
Galilean and the Poisson algebra. When the stability principle is applied to
the Poincar\'{e}-Heisenberg algebra, two deformation parameters emerge which
define two length (or time) scales. In addition there are, for each of them,
a plus or minus sign possibility in the relevant commutators. One of the
deformation length scales, related to non-commutativity of momenta, is
probably related to the Planck length scale but the other might be much
larger. In this paper this is used as a working hypothesis to look for
physical effects that might settle this question. Phase-space modifications,
deviations from $c$ in speed measurements of massless wave packets,
resonances, interference, electron spin resonance and non-commutative QED
are considered.
\end{abstract}


\section{Introduction}

The idea of modifying the algebra of the space-time components $x_{\mu }$ in
such a way that they become non-commuting operators has appeared many times
in the physical literature (\cite{Snyder} \cite{Yang} \cite{Kadyshevsky1} 
\cite{Kadyshevsky2} \cite{Banai}, \cite{Das} \cite{Atkinson} \cite{Gudder} 
\cite{Finkelstein} \cite{Dineykhan} \cite{Lee} \cite{Prugovecki} \cite%
{Blokhintsev} \cite{Schild} \cite{Veneziano} \cite{t Hooft} \cite{Jackson} 
\cite{Madore} \cite{Maggiore} \cite{Kempf}, etc.). The aim of most of these
proposals was to endow space-time with a discrete structure, to be able, for
example, to construct quantum fields free of ultraviolet divergences.
Sometimes a non-zero commutator is simply postulated, in some other
instances the motivation is the formulation of field theory in curved
spaces. String theories \cite{string} \cite{JHEP} and quantum relativity 
\cite{quantrelat} \cite{Dubois} have also provided hints concerning the
non-commutativity of space-time at a fundamental level.

A somewhat different point of view has been proposed in \cite{Vilela1} \cite%
{Vilela3}. There the space-time noncommutative structure is arrived at
through the application of the \textit{stability of physical theories
principle} (SPT). The rationale behind this principle is the fact that the
parameters entering in physical theories are never known with absolute
precision. Therefore, robust physical laws with a wide range of validity can
only be those that do not change in a qualitative manner under a small
change of parameters, that is, \textit{stable} (or \textit{rigid}) theories.
The stable-model point of view originated in the field of non-linear
dynamics, where it led to the notion of \textit{structural stability }\cite%
{Andronov} \cite{Smale}. Later on, Flato \cite{Flato} and Faddeev \cite%
{Faddeev} have shown that the same pattern occurs in the fundamental
theories of Nature, namely the transition from non-relativistic to
relativistic and from classical to quantum mechanics, may be interpreted as
the replacement of two unstable theories by two stable ones. The stabilizing
deformations lead, in the first case, from the Galilean to the Lorentz
algebra and, in the second one, from the algebra of commutative phase-space
to the Moyal-Vey algebra (or equivalently to the Heisenberg algebra). The
deformation parameters are $\frac{1}{c}$ (the inverse of the speed of light)
and $h$ (the Planck constant). Except for the isolated zero value, the
deformed algebras are all equivalent for non-zero values of $\frac{1}{c}$
and $h$. Hence, relativistic mechanics and quantum mechanics might have been
derived from the conditions for stability of two mathematical structures,
although the exact values of the deformation parameters cannot be fixed by
purely algebraic considerations. Instead, the deformation parameters are
fundamental constants to be obtained from experiment and, in this sense, not
only is deformation theory the theory of stable theories, it is also the
theory that identifies the fundamental constants.

The SPT principle is related to the idea that physical theories drift
towards simple algebras \cite{Segal} \cite{Finkelstein2} \cite{Finkelstein3}%
, because all simple algebras are stable, although not all stable algebras
are simple.

When the SPT principle is applied to the algebra of relativistic quantum
mechanics (the Poincar\'{e}-Heisenberg algebra)%
\begin{equation}
\begin{array}{rcl}
\lbrack M_{\mu \nu },M_{\rho \sigma }] & = & i(M_{\mu \sigma }\eta _{\nu
\rho }+M_{\nu \rho }\eta _{\mu \sigma }-M_{\nu \sigma }\eta _{\mu \rho
}-M_{\mu \rho }\eta _{\nu \sigma }) \\ 
\lbrack M_{\mu \nu },p_{\lambda }] & = & i(p_{\mu }\eta _{\nu \lambda
}-p_{\nu }\eta _{\mu \lambda }) \\ 
\lbrack M_{\mu \nu },x_{\lambda }] & = & i(x_{\mu }\eta _{\nu \lambda
}-x_{\nu }\eta _{\mu \lambda }) \\ 
\lbrack p_{\mu },p_{\nu }] & = & 0 \\ 
\lbrack x_{\mu },x_{\nu }] & = & 0 \\ 
\lbrack p_{\mu },x_{\nu }] & = & i\eta _{\mu \nu }\boldsymbol{1}%
\end{array}
\label{I.1}
\end{equation}%
$\eta _{\mu \nu }=(1,-1,-1,-1)$, $c=\hbar =1$, it leads \cite{Vilela1} to%
\begin{equation}
\begin{array}{rcl}
\lbrack M_{\mu \nu },M_{\rho \sigma }] & = & i(M_{\mu \sigma }\eta _{\nu
\rho }+M_{\nu \rho }\eta _{\mu \sigma }-M_{\nu \sigma }\eta _{\mu \rho
}-M_{\mu \rho }\eta _{\nu \sigma }) \\ 
\lbrack M_{\mu \nu },p_{\lambda }] & = & i(p_{\mu }\eta _{\nu \lambda
}-p_{\nu }\eta _{\mu \lambda }) \\ 
\lbrack M_{\mu \nu },x_{\lambda }] & = & i(x_{\mu }\eta _{\nu \lambda
}-x_{\nu }\eta _{\mu \lambda }) \\ 
\lbrack p_{\mu },p_{\nu }] & = & -i\frac{\epsilon ^{^{\prime }}}{R^{2}}%
M_{\mu \nu } \\ 
\lbrack x_{\mu },x_{\nu }] & = & -i\epsilon \ell ^{2}M_{\mu \nu } \\ 
\lbrack p_{\mu },x_{\nu }] & = & i\eta _{\mu \nu }\Im  \\ 
\lbrack p_{\mu },\Im ] & = & -i\frac{\epsilon ^{^{\prime }}}{R^{2}}x_{\mu }
\\ 
\lbrack x_{\mu },\Im ] & = & i\epsilon \ell ^{2}p_{\mu } \\ 
\lbrack M_{\mu \nu },\Im ] & = & 0%
\end{array}
\label{I.2}
\end{equation}%
The stabilization of the Poincar\'{e}-Heisenberg algebra has been further
studied and extended in \cite{Chrysso} \cite{Ahluwalia1} \cite{Ahluwalia2}.
The essential message from (\ref{I.2}) or from the slightly more general
form obtained in \cite{Chrysso} is that from the unstable Poincar\'{e}%
-Heisenberg algebra $\{M_{\mu \nu },p_{\mu },x_{\nu }\}$ one obtains a
stable algebra with two deformation parameters $\ell $ and $\frac{1}{R}$. In
addition there are two undetermined signs $\epsilon $ and $\epsilon ^{\prime
}$and the central element of the Heisenberg algebra becomes a non-trivial
operator $\Im $. The existence of two continuous deformation parameters when
the algebra is stabilized is a novel feature of the deformation point of
view, which does not appear in other noncommutative space-time approaches.
These deformation parameters may define two different length scales. Of
course, once one of them is identified as a fundamental constant, the other
will be a pure number.

Being associated to the noncommutativity of the generators of space-time
translations, the parameter $\frac{1}{R}$ may be associated to space-time
curvature and therefore might not be relevant for considerations related to
the tangent space. It is, of course, very relevant for quantum gravity
studies \cite{Ahluwalia2}. Already in the past, some authors \cite{Faddeev},
have associated the noncommutativity of translations to gravitational
effects, the gravitation constant being the deformation parameter.
Presumably then $\frac{1}{R}$ might be associated to the Planck length
scale. However $\ell $, the other deformation parameter, defines a
completely independent length scale which might be much closer to laboratory
phenomena. This will be the working hypothesis to be explored in this paper.
Therefore when $\frac{1}{R}$ is assumed to be very small the deformed
algebra may be approximated by%
\begin{equation}
\begin{array}{rcl}
\lbrack M_{\mu \nu },M_{\rho \sigma }] & = & i(M_{\mu \sigma }\eta _{\nu
\rho }+M_{\nu \rho }\eta _{\mu \sigma }-M_{\nu \sigma }\eta _{\mu \rho
}-M_{\mu \rho }\eta _{\nu \sigma }) \\ 
\lbrack M_{\mu \nu },p_{\lambda }] & = & i(p_{\mu }\eta _{\nu \lambda
}-p_{\nu }\eta _{\mu \lambda }) \\ 
\lbrack M_{\mu \nu },x_{\lambda }] & = & i(x_{\mu }\eta _{\nu \lambda
}-x_{\nu }\eta _{\mu \lambda }) \\ 
\lbrack p_{\mu },p_{\nu }] & = & 0 \\ 
\lbrack x_{\mu },x_{\nu }] & = & -i\epsilon \ell ^{2}M_{\mu \nu } \\ 
\lbrack p_{\mu },x_{\nu }] & = & i\eta _{\mu \nu }\Im \\ 
\lbrack p_{\mu },\Im ] & = & 0 \\ 
\lbrack x_{\mu },\Im ] & = & i\epsilon \ell ^{2}p_{\mu } \\ 
\lbrack M_{\mu \nu },\Im ] & = & 0%
\end{array}
\label{I.3}
\end{equation}%
For future reference this algebra will be denoted $\mathcal{R}_{\ell ,\infty
}$. Notice that in relation to the more general deformation obtained in \cite%
{Chrysso}, we are also considering $\alpha _{3}=0$ (or $\beta =0$ in \cite%
{Ahluwalia2}). The nature of the sign $\epsilon $ has physical consequences.
If $\epsilon =+1$ time will have a discrete spectrum, whereas if $\epsilon
=-1$ it is when one the space coordinates is diagonalized that discrete
spectrum is obtained. In this sense if $\epsilon =+1$, $\ell $ might be
called "the fundamental time" and "the fundamental length" if $\epsilon =-1$%
. In this paper one discusses consequences of both signs.

General (noncommutative) geometry properties of the algebra (\ref{I.3}) have
been studied before \cite{Vilela3} as well as some other consequences \cite%
{Vilela2} \cite{Carlen} \cite{Vilela4} \cite{Dzhunu} \cite{Goldin}. Here the
emphasis will be on effects which might be detectable at the laboratory
level, if the working hypothesis that $\ell $ defines a much larger scale
than Planck's is true. In addition, some of the non-commutativity and
time-discreteness effects that have been proposed in the past will be
discussed, in particular to find out whether they are or not relevant as a
test of the algebra (\ref{I.3}).

In the recent past, most papers dealing with space-time non-commutativity
start from the hypothesis%
\begin{equation}
\lbrack x_{\mu },x_{\nu }]=i\theta _{\mu \nu }  \label{I.4}
\end{equation}%
$\theta _{\mu \nu }$ being a c-number antisymmetric tensor (for a review see 
\cite{Hinchliffe1}). Then, calculations are carried out by replacing the
usual product of functions in space-time by the $\ast -$product%
\begin{equation}
\left( f\ast g\right) \left( x\right) =f\left( x\right) e^{i\overset{%
\longleftarrow }{\partial ^{\mu }}\theta _{\mu \nu }\overset{\longrightarrow 
}{\partial ^{\nu }}}g\left( x\right)  \label{I.5}
\end{equation}%
A similar $\ast -$product formulation may be implemented for the algebra (%
\ref{I.2}) by replacing in (\ref{I.5}) $\theta _{\mu \nu }$ by the operator $%
M_{\mu \nu }$ and, to have a full $\ast -$product formulation, using the
Moyal product for products of functions of coordinates and momenta. However
the situation is quite different from the one implied by (\ref{I.4}) because 
$M_{\mu \nu }$ is an operator, not a constant tensor deformation parameter.
Hence it does not lead to Lorentz violation, the deformed algebras (\ref{I.2}%
-\ref{I.3}) being consistent with preservation of Lorentz invariance.
Therefore some of the tests proposed for (\ref{I.4}) are not relevant for (%
\ref{I.3}). In addition the deformed algebras introduce a new non-trivial
operator $\Im $ which replaces the central element of the Heisenberg
algebra. In particular this operator corresponds to an additional component
in the most general connections compatible with (\ref{I.3}) \cite{Vilela3}.

When the right-hand side of (\ref{I.4}) is a c-number $\theta _{\mu \nu }$,
with dimensions of length-squared, it may be roughly interpreted as the
smallest patch of area in the $\mu \nu -$plane that one may consider to be
observable, like $\hslash $ in $[x_{i},p_{j}]=i\hslash \delta _{ij}$ may be
interpreted as the smallest patch in phase space. However here the
right-hand-side of the commutators is an operator and the interpretation is
subtler.

The present paper is concerned with the discussion of effects which might
lead to actual experimental tests if $\ell $ is not too small. It must be
pointed out that some of these effects, as it will be referred to in the
appropriate places, may have already been suggested by other authors.
Nevertheless in most cases they are suggested in the framework of a simple
time or space discreteness hypothesis, without the benefit of a full
space-time algebra. For that reason some of the conclusions are different or
more detailed.

Phase-space modifications, deviations from $c$ in speed measurements of
massless wave packets, resonances, interference, electron spin resonance and
non-commutative QED are considered. Finally, in the Appendix, some explicit
representations of the space-time algebra are collected, which are useful
for the calculations.

\section{Phase space effects}

\subsection{Cross-sections and particle multiplicity}

If $\epsilon =+1$ there is a phase space contraction effect at high
energies. This was discussed in \cite{Vilela4}, being pointed out that it
might relevant for the calculation of the GZK radius \cite{Greisen} \cite%
{Zatsepin}. From the calculations in \cite{Vilela4}, the conclusion was
that, whereas the value of the GZK cutoff would not be much changed, the
radius of the GZK sphere would increase, allowing for more nucleons from
farther distances to reach earth at energies above $5.10^{9}$eV. For this
effect to be detectable the fundamental time should not be smaller than $%
10^{-26}$ seconds.

Here, further consequences of the phase-space modification are studied. Both
signs $\epsilon =+1$ and $\epsilon =-1$ are considered. In particular, if
the conjecture that the scale $\ell $ is much larger than the Planck's scale
is true, such effects might already be observed at the energy of the
existing colliders.

The modification of the density of states \cite{Vilela4} is obtained by
computing how many available states a particle of momentum $p$ has, for
example, in a scattering experiment. Once the direction of $p$ is fixed, the
problem becomes a one-dimensional problem, which may be dealt with by a
subalgebra $\left\{ x^{1},p^{1},\Im \right\} $ of (\ref{I.3}). Let $\epsilon
=+1$ and define hyperbolic coordinates in the plane $\left( p^{1},\Im
\right) $%
\begin{equation}
\begin{array}{lll}
p^{1} & = & \frac{r}{\ell }\sinh \mu \\ 
\Im & = & r\cosh \mu%
\end{array}
\label{ps01}
\end{equation}%
Then%
\begin{equation*}
\frac{\partial }{\partial \mu }=\frac{1}{\ell }\Im \frac{\partial }{\partial
p^{1}}+\ell p^{1}\frac{\partial }{\partial \Im }
\end{equation*}%
one comparing with the representation (\ref{A.8}) one obtains%
\begin{equation}
x^{1}=i\ell \frac{\partial }{\partial \mu }  \label{ps02}
\end{equation}%
or, equivalently%
\begin{equation}
\begin{array}{lll}
x^{1} & = & x \\ 
p^{1} & = & \frac{r}{\ell }\sinh \left( \frac{\ell }{i}\frac{\partial }{%
\partial x}\right) \\ 
\Im & = & r\cosh \left( \frac{\ell }{i}\frac{\partial }{\partial x}\right)%
\end{array}
\label{ps03}
\end{equation}%
For $r=1$ and $\ell \rightarrow 0$, the classical result is obtained.
Hereafter let us consider $r=1$. In the $x-$basis the eigenvectors $p$ of
the momentum $p^{1}$ are $e^{ik_{n}x}$ which, with vanishing boundary
conditions on a box, has eigenvalues 
\begin{equation}
p_{n}=\frac{1}{\ell }\sinh \left( \frac{\pi }{L}n\ell \right)  \label{ps1}
\end{equation}%
corresponding to $k_{n}=\frac{\pi n}{L}$. Therefore the number of states
with momenta smaller than $p$ is%
\begin{equation}
G_{+}^{1D}\left( p\right) =\frac{L}{\pi }\frac{1}{\ell }\sinh ^{-1}\left(
\ell p\right)  \label{ps2}
\end{equation}%
and the density of states is%
\begin{equation}
g_{+}^{1D}\left( p\right) =\frac{dG_{+}^{1D}}{dp}=\frac{L}{\pi }\frac{1}{%
\sqrt{1+\ell ^{2}p^{2}}}  \label{ps3}
\end{equation}%
For three dimensions, considering the number of independent states with
absolute momentum less than $\left\vert p\right\vert $%
\begin{equation}
G_{+}^{3D}\left( \left\vert p\right\vert \right) =\frac{V}{6\pi ^{2}}\frac{1%
}{\ell ^{3}}\left( \sinh ^{-1}\left( \ell \left\vert p\right\vert \right)
\right) ^{3}  \label{ps4}
\end{equation}%
leads to a density of states%
\begin{equation}
g_{+}^{3D}\left( \left\vert p\right\vert \right) =\frac{V}{2\pi ^{2}}\frac{1%
}{\ell ^{2}}\frac{\left( \sinh ^{-1}\left( \ell \left\vert p\right\vert
\right) \right) ^{2}}{\sqrt{1+\ell ^{2}\left\vert p\right\vert ^{2}}}
\label{ps5}
\end{equation}

For $\epsilon =-1$ the appropriate coordinates in the plane $\left(
p^{1},\Im \right) $ are $p^{1}=\frac{r}{\ell }\sin \theta $, $\Im =r\cos
\theta $, $x^{1}=i\ell \frac{\partial }{\partial \theta }$ and one would
obtain the opposite effect, namely the factor $1/\sqrt{1-\ell ^{2}p^{2}}$).
Therefore%
\begin{equation}
g_{-}^{1D}\left( p\right) =\frac{L}{\pi }\frac{1}{\sqrt{1-\ell ^{2}p^{2}}}
\label{ps3a}
\end{equation}%
\begin{equation}
g_{-}^{3D}\left( \left\vert p\right\vert \right) =\frac{V}{2\pi ^{2}}\frac{1%
}{\ell ^{2}}\frac{\left( \sin ^{-1}\left( \ell \left\vert p\right\vert
\right) \right) ^{2}}{\sqrt{1-\ell ^{2}\left\vert p\right\vert ^{2}}}
\label{ps5a}
\end{equation}

\begin{figure}[htb]
\begin{center}
\psfig{figure=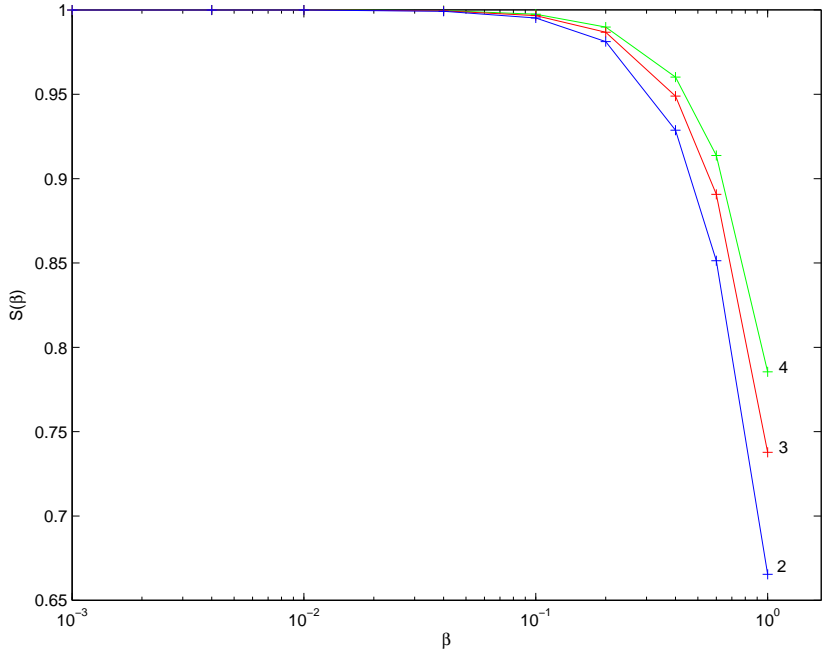,width=11truecm}
\end{center}
\caption{Phase-space suppression
function ($\epsilon =+1$) for 2,3 and 4 final particles}
\label{supress}
\end{figure}

The conclusion is that for $\epsilon =+1$ there is a contraction of phase
space increasing with energy and an expansion for $\epsilon =-1$, the cross
sections being corrected by the new density of states (\ref{ps5}) and (\ref%
{ps5a}). For $\epsilon =+1$ the suppression effect of the phase-space
contraction on high energy reactions may be estimated by comparing the
integral 
\begin{equation*}
I_{N}\left( \ell \right) =\int \cdots \int_{0}^{\omega }\frac{\left( \sinh
^{-1}\left( \ell p_{1}\right) \right) ^{2}dp_{1}}{\ell ^{2}\sqrt{1+\ell
^{2}p_{1}^{2}}}\cdots \frac{\left( \sinh ^{-1}\left( \ell p_{N}\right)
\right) ^{2}dp_{N}}{\ell ^{2}\sqrt{1+\ell ^{2}p_{N}^{2}}}\max \left( \omega
-\sum_{i=1}^{N}p_{i},0\right) 
\end{equation*}%
with $I_{N}\left( 0\right) $. This estimates the suppression effect on an
high energy final state integral for total energy $\omega $ neglecting
masses. Changing variables one obtains 
\begin{equation*}
I_{N}\left( \ell \right) =\omega ^{3}\int_{0}^{1}\frac{\left( \sinh
^{-1}\left( \beta x_{1}\right) \right) ^{2}dx_{1}}{\beta ^{2}\sqrt{1+\beta
^{2}x_{1}^{2}}}\cdots \frac{\left( \sinh ^{-1}\left( \beta x_{N}\right)
\right) ^{2}dx_{N}}{\beta ^{2}\sqrt{1+\beta ^{2}x_{N}^{2}}}\max \left(
1-\sum_{i=1}^{N}x_{i},0\right) 
\end{equation*}%
with $\beta =\omega \ell $. Fig.\ref{supress} is a plot of the suppression
function $S\left( \beta \right) =\frac{I_{N}\left( \ell \right) }{%
I_{N}\left( 0\right) }$ for $N=2,3,4$. One sees that the suppression effect
decreases when the number of final particles increases. The phase space
suppression effect implies that if cross section values found at low
energies are used to predict the final states at higher energies, an
increase in particle multiplicity will be found above the expected one.
 
For $\epsilon =-1$ the effect would
be the opposite one, that is, a smaller multiplicity. One also sees that
these effects will only become noticeable for $\beta \sim O\left( 1\right) $%
. For example, an observation of the effects starting at around $\omega =600$
Gev would imply $\ell \backsim O\left( 10^{-27}s\right) $ or $\ell \backsim
O\left( 3\times 10^{-17}cm\right) $.

\subsection{The degeneracy pressure}

That the phase space volume modifications at high energies (contraction for $%
\epsilon =+1$, dilatation for $\epsilon =-1$) would also lead to statistical
mechanics predictions was briefly mentioned in \cite{Carlen}. This might, in
particular, have some consequences for models of dense star matter. Here I
will analyze the modifications implied by the deformed algebra on the
degeneracy pressure of a Fermi gas. Both the non-relativistic and the
relativistic case will be analyzed. For a gas of nonrelativistic particles
the kinetic energy $E$ is%
\begin{equation*}
E=\frac{p^{2}}{2m}
\end{equation*}%
For $\epsilon =+1$, changing variables in (\ref{ps4}) the density of states
becomes%
\begin{equation*}
g_{+}^{3D}\left( E\right) =\frac{V}{4\pi ^{2}}\frac{1}{\ell ^{2}}\left(
\sinh ^{-1}\left( \ell \sqrt{2mE}\right) \right) ^{2}\frac{\sqrt{2m}}{\sqrt{E%
}\sqrt{1+2\ell ^{2}mE}}
\end{equation*}%
From%
\begin{equation*}
N=\int_{0}^{E_{F}}\left( 2s+1\right) g_{+}^{3D}\left( E\right) dE
\end{equation*}%
at $T=0$, one obtains%
\begin{equation*}
E_{F}=\frac{1}{2m\ell ^{2}}\sinh \left( \frac{6\pi ^{2}\ell ^{3}N}{\left(
2s+1\right) V}\right) ^{2/3}
\end{equation*}%
In leading $\ell ^{2}$ order the energy and the pressure are%
\begin{eqnarray*}
U\left( 0\right)  &=&\int_{0}^{E_{F}}\left( 2s+1\right) Eg_{+}^{3D}\left(
E\right) d\varepsilon  \\
&\simeq &\left( 2s+1\right) \frac{V}{4\pi ^{2}}\left( 2m\right)
^{3/2}\left\{ \frac{2}{5}E_{F}^{5/2}-\frac{10}{7}m\ell ^{2}E^{7/2}\right\} 
\end{eqnarray*}%
\begin{eqnarray*}
P &=&-\left( \frac{\partial U\left( 0\right) }{\partial V}\right) _{N,S} \\
&\simeq &\frac{\left( 2s+1\right) }{4\pi ^{2}}\left( 2m\right) ^{3/2}\left\{ 
\frac{4}{15}E_{F}^{5/2}-\frac{4}{7}m\ell ^{2}E^{7/2}\right\} 
\end{eqnarray*}%
leading to%
\begin{equation*}
P\simeq \frac{2}{3}\frac{U\left( 0\right) }{V}+\frac{15}{14}\left( \frac{%
20\pi ^{2}m}{2s+1}\right) ^{2/5}\ell ^{2}\left( \frac{U\left( 0\right) }{V}%
\right) ^{7/5}
\end{equation*}

In the relativistic case, which is the one that is relevant, for example,
for neutron star matter, the total energy density per unit volume is%
\begin{equation*}
\rho =\int_{0}^{p_{F}}\left( p^{2}+m^{2}\right) ^{1/2}\frac{1}{V}%
g_{+}^{3D}\left( p\right) dp
\end{equation*}%
and the pressure%
\begin{equation*}
P=\frac{1}{3}\int_{0}^{p_{F}}\frac{p^{2}}{\left( p^{2}+m^{2}\right) ^{1/2}}%
\frac{1}{V}g_{+}^{3D}\left( p\right) dp
\end{equation*}%
with $p_{F}=\frac{1}{\ell }\sinh \left( \frac{6\pi ^{2}\ell ^{3}N}{\left(
2s+1\right) V}\right) ^{1/3}$. Writing $\rho $ and $P$ with the adimensional
variables $m\ell $ and $\frac{N}{Vm^{3}}=\frac{n}{m^{3}}$%
\begin{equation*}
\rho _{\ell }\left( \ell m,\frac{n}{m^{3}}\right) =\frac{\left( 2s+1\right)
m^{4}}{2\pi ^{2}}\int_{0}^{p_{F}/m}\frac{dx}{\left( \ell m\right) ^{2}}%
\left( \sinh ^{-1}\left( \ell mx\right) \right) ^{2}\frac{\sqrt{1+x^{2}}}{%
\sqrt{1+\ell ^{2}m^{2}x^{2}}}
\end{equation*}%
\begin{equation*}
P_{\ell }\left( \ell m,\frac{n}{m^{3}}\right) =\frac{\left( 2s+1\right) m^{4}%
}{6\pi ^{2}}\int_{0}^{p_{F}/m}\frac{dx}{\left( \ell m\right) ^{2}}\left(
\sinh ^{-1}\left( \ell mx\right) \right) ^{2}\frac{x^{2}}{\sqrt{\left(
1+x^{2}\right) \left( 1+\ell ^{2}m^{2}x^{2}\right) }}
\end{equation*}%

\begin{figure}[htb]
\begin{center}
\psfig{figure=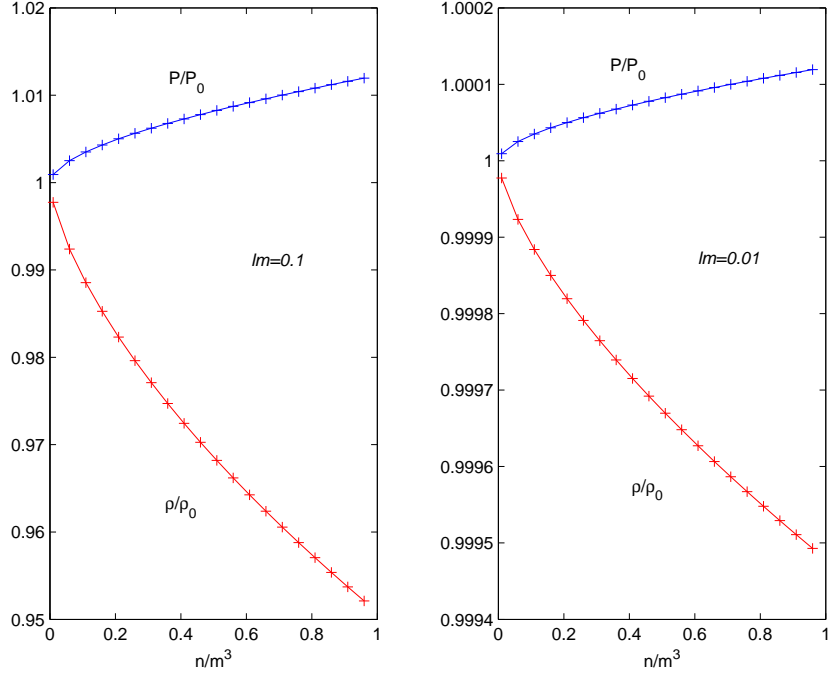,width=11truecm}
\end{center}
\caption{Ratios $\rho _{\ell }/\rho _{0}$ and 
$P_{\ell }/P_{0}$ for two $\ell m$ values}
\label{pressure}
\end{figure}

In the Fig.\ref{pressure} the ratios $\rho _{\ell }/\rho _{0}$ and 
$P_{\ell }/P_{0}$ are plotted for two $\ell m$ values. One sees that the
phase-space suppression ($\epsilon =+1$) implies a larger degeneracy
pressure and a smaller total energy density for the same $n=\frac{N}{V}$.
However, (in contrast with the effects that might seen at high energy
colliders, as discussed above) and for reasonable star matter densities the
effect is probably too small to be observed. For example, with $m$ the
neutron mass, a matter density of $4.10^{14}g/cm^{3}$ and $\ell =10^{-26}s$
one has $\ell m=1.35\times 10^{-2}$ but only $\frac{n}{m^{3}}=2.16\times
10^{-3}$. One sees from Fig.\ref{pressure} that for these values the effect
is extremely small.

The corresponding results for the case $\epsilon =-1$ are obtained by
replacing $\left( 1+\ell ^{2}m^{2}x^{2}\right) $ by $\left( 1-\ell
^{2}m^{2}x^{2}\right) $ and $\sinh ^{-1}$ by $\sin ^{-1}$ in the equations
above.

\section{Measuring the speed of wave-packets}

In the noncommutative context, because the space and the time coordinates
cannot be simultaneously diagonalized, speed can only be defined in terms of
expectation values, for example%
\begin{equation}
v_{\psi }^{i}=\frac{1}{\left\langle \psi _{t},\psi _{t}\right\rangle }\frac{d%
}{dt}\left\langle \psi _{t},x^{i}\psi _{t}\right\rangle  \label{sw1}
\end{equation}%
Here, one considers a normalized state $\psi $ with a small dispersion of
momentum around a central value $k$. At time zero%
\begin{equation}
\psi _{0}=\int \left\vert p^{0}\overset{\longrightarrow }{p}\alpha
\right\rangle f_{k}\left( p\right) d^{3}p  \label{sw2}
\end{equation}%
where $p^{0}=\sqrt{\left\vert \overset{\longrightarrow }{p}\right\vert
^{2}+m^{2}}$, $\alpha $ standing for the quantum numbers associated to the
little group of $p$ and $f_{k}\left( p\right) $ a normalized function peaked
around $p=k$.

To obtain $\psi _{t}$ one should apply to $\psi _{0}$ the time-shift
operator. However this is not $p^{0}$ because%
\begin{equation}
e^{-i\alpha p^{0}}te^{i\alpha p^{0}}=t+\alpha \Im  \label{sw3}
\end{equation}%
follows from 
\begin{equation}
\left[ p^{0},t\right] =i\Im  \label{sw4}
\end{equation}%
whereas a time-shift generator $\Gamma $ should satisfy%
\begin{equation}
\left[ \Gamma ,t\right] =i\mathbf{1}  \label{sw5}
\end{equation}%
In order $O\left( \ell ^{4}\right) $ one may take%
\begin{equation}
\Gamma =p^{0}\Im ^{-1}-\frac{\epsilon }{3}\ell ^{2}\left( p^{0}\right)
^{3}\Im ^{-3}  \label{sw6}
\end{equation}%
because%
\begin{equation}
\left[ \Gamma ,t\right] =i\left( 1-\ell ^{4}\left( p^{0}\right) ^{4}\Im
^{-4}\right)  \label{sw7}
\end{equation}%
To obtain this result, use was made of $\left[ t,\Im ^{-1}\right]
=-i\epsilon \ell ^{2}p^{0}\Im ^{-2}$, which follows from $\left[ t,\Im \Im
^{-1}\right] =0$.

Here one uses a basis where the set $\left( p^{\mu },\Im \right) $ is
diagonalized and define%
\begin{equation}
\overset{\thicksim }{p^{\mu }}=\frac{p^{\mu }}{\Im }  \label{sw8}
\end{equation}%
$\overset{\thicksim }{p^{\mu }}$ is the momentum in units of $\Im $.

Therefore up to the same $O\left( \ell ^{4}\right) $ order%
\begin{equation}
\psi _{t}=\int e^{-i\left( \overset{\thicksim }{p^{0}}-\frac{\epsilon }{3}%
\ell ^{2}\left( \overset{\thicksim }{p^{0}}\right) ^{3}\right) }\left\vert 
\overset{\thicksim }{p^{0}}\overset{\thicksim }{p^{i}}\alpha \right\rangle
f_{k}\left( \overset{\thicksim }{p}\right) d^{3}\overset{\thicksim }{p}
\label{sw9}
\end{equation}

To compute the expectation value of $x^{i}$ one notices that from%
\begin{equation}
x^{\mu }=i\left( \epsilon \ell ^{2}p^{\mu }\frac{\partial }{\partial \Im }%
-\Im \frac{\partial }{\partial p_{\mu }}\right)   \label{sw10}
\end{equation}%
using $\frac{\partial }{\partial \Im }=-\frac{p^{\nu }}{\Im ^{2}}\frac{%
\partial }{\partial \overset{\thicksim }{p^{\nu }}}$ one obtains%
\begin{equation}
x^{\mu }=-i\left( \frac{\partial }{\partial \overset{\thicksim }{p_{\mu }}}%
+\epsilon \ell ^{2}\left\{ \overset{\thicksim }{p^{\mu }}\overset{\thicksim }%
{p^{\nu }}\frac{\partial }{\partial \overset{\thicksim }{p^{\nu }}}\right\}
_{S}\right)   \label{sw11}
\end{equation}%
$\left\{ {}\right\} _{S}$ meaning symmetrization of the operators.

Now the expectation value of this operator in the state $\psi _{t}$ is
computed and taking the time derivative one obtains for the wave packet
speed in order $\ell ^{2}$%
\begin{equation}
v_{\psi }=\frac{\overset{\thicksim }{p}}{\overset{\thicksim }{p^{0}}}\left(
1-\epsilon \ell ^{2}\left( p^{0}\right) ^{2}\right) -\epsilon \ell
^{2}\left( \overset{\thicksim }{p}\overset{\thicksim }{p^{0}+\left( \overset{%
\thicksim }{p}\right) ^{2}}\frac{\overset{\thicksim }{p}}{\overset{\thicksim 
}{p^{0}}}\right)   \label{sw12}
\end{equation}%
$\ell ^{2}$ being small, this deviation from $\frac{\overset{\thicksim }{p}}{%
\overset{\thicksim }{p^{0}}}$ may be difficult to detect for massive
particles given the uncertainty on the values of the mass. However, for
massless particles the deviation from $c\left( =1\right) $%
\begin{equation}
\Delta v_{\psi }=-3\epsilon \ell ^{2}\left( p^{0}\right) ^{2}  \label{sw13}
\end{equation}%
might already be possible to detect accurately with present experimental
means \cite{opera}. Such deviation above or below the speed $c$ (depending
on the sign of $\epsilon $) would not imply any modification of the
relativistic deformation constant $\left( \frac{1}{c}\right) $, nor a
breakdown of relativity. Rather, it would be a manifestation of the
noncommutative space-time structure.

For $p^{0}=20$ GeV and $\ell =3\times 10^{-18}cm$ (or $\ell =10^{-28}s$) $%
\left\vert \Delta v_{\psi }\right\vert \sim 2.7\times 10^{-5}$.

\section{Time quantization and resonances}

Some years ago Ehrlich \cite{Ehrlich1}, finding a regularity in the
resonance widths known at the time, conjectured that the widths might be
quantized in multiples of some fundamental time unit. Later, the same author 
\cite{Ehrlich2} using more recent data, pointed out that the quantization
hypothesis of the resonance widths did not agree as well as before.
Nevertheless, the conjecture has its merit and deserves to be checked within
the present framework. If the time quantization has a direct bearing on
resonance widths it should already be apparent in simple potential models.
Consider the simple one dimensional potential displayed in Fig.\ref%
{potential}.

\begin{figure}[htb]
\begin{center}
\psfig{figure=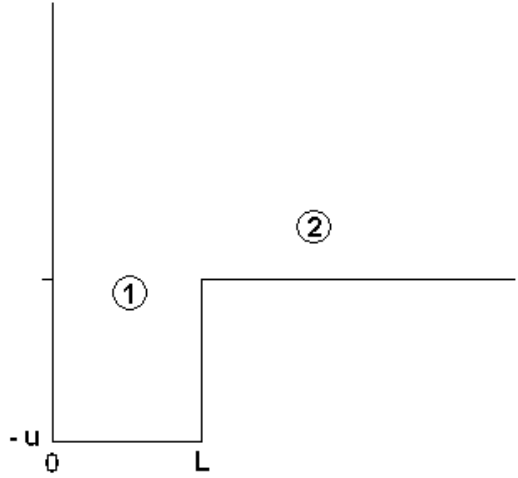,width=11truecm}
\end{center}
\caption{A simple one-dimensional potential}
\label{potential}
\end{figure}

Let the
wave functions in regions 1 and 2 be 
\begin{equation}
\begin{array}{lll}
\Psi ^{(1)} & = & A\sin \left( kx\right) \\ 
\Psi ^{(2)} & = & e^{-ik^{(2)}x}+Be^{ik^{(2)}x}%
\end{array}
\label{2.1}
\end{equation}%
Quantized time would correspond to $\epsilon =+1$, therefore, using the
representation (\ref{ps03}), 
\begin{equation}
\begin{array}{lll}
x & = & x \\ 
p & = & -i\frac{r}{\ell }\sin \left( \ell \frac{d}{dx}\right) \\ 
\mathcal{\Im } & = & r\cos \left( \ell \frac{d}{dx}\right)%
\end{array}
\label{2.1a}
\end{equation}%
the momentum $p$ associated to a wave number $k$ is 
\begin{equation}
p=\frac{r}{\ell }\sinh \left( k\ell \right)  \label{2.1b}
\end{equation}%
Using the matching conditions at $x=L$ one obtains 
\begin{equation}
A=\frac{2e^{-ik^{(2)}L}}{\sin \left( kL\right) +\frac{ik}{k^{(2)}}\cos
\left( kL\right) }  \label{2.2}
\end{equation}%
and 
\begin{equation}
B=e^{-i2k^{(2)}L}\left\{ \frac{2\sin \left( kL\right) }{\sin \left(
kL\right) +\frac{ik}{k^{(2)}}\cos \left( kL\right) }-1\right\}  \label{2.3}
\end{equation}%
with $k^{(2)}$ obtained from $k$ by the matching of the energy in regions 1
and 2. For a non-relativistic approximation and $r=1$ it is 
\begin{equation}
\cosh \left( 2k^{(2)}\ell \right) =\cosh \left( 2k\ell \right) -4mu\ell ^{2}
\label{2.4}
\end{equation}%
and in the relativistic case%
\begin{equation}
\cosh \left( 2k^{(2)}\ell \right) =1+\left( \sqrt{\cosh \left( 2k\ell
\right) -1+2m_{0}^{2}\ell ^{2}}-\sqrt{2}\ell u\right) ^{2}-2m_{0}^{2}\ell
^{2}  \label{2.5}
\end{equation}

\begin{figure}[htb]
\begin{center}
\psfig{figure=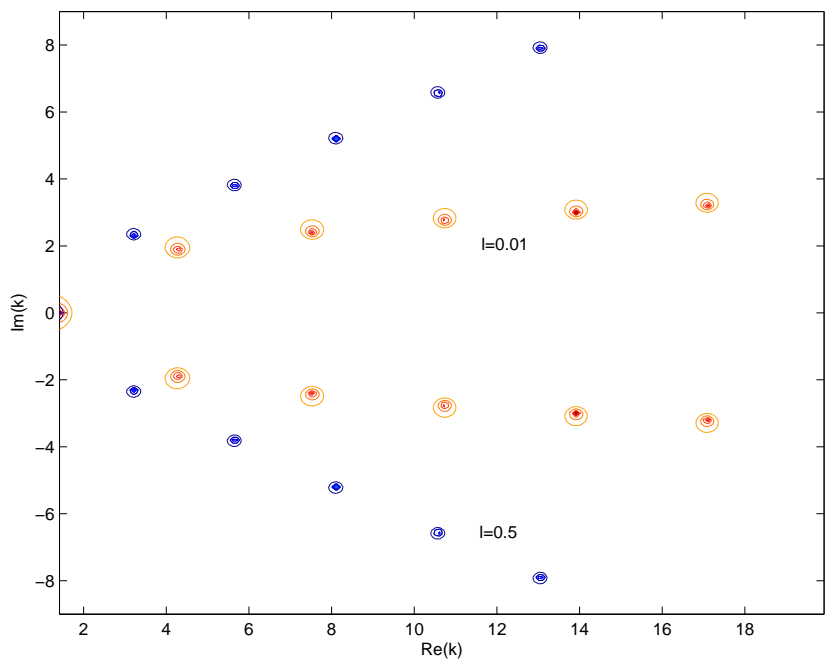,width=11truecm}
\end{center}
\caption{The complex zeros of the
function $F\left( k\right) $}
\label{zeros}
\end{figure}

The resonances are associated to the complex zeros of $\sin \left( kL\right)
+\frac{ik}{k^{(2)}}\cos \left( kL\right) $, that is using (\ref{2.4}), to
the zeros of the function%
\begin{equation*}
F\left( k\right) =\cosh \left( 2k\ell \right) -4mu\ell ^{2}-\cos \left( 
\frac{2k\ell }{\tan \left( kL\right) }\right)
\end{equation*}%
In Fig.\ref{zeros} the location of the zeros of $F\left( k\right) $ are
plotted for $m=u=L=1$, $\ell =0.01$ and $\ell =0.5$.

\begin{figure}[htb]
\begin{center}
\psfig{figure=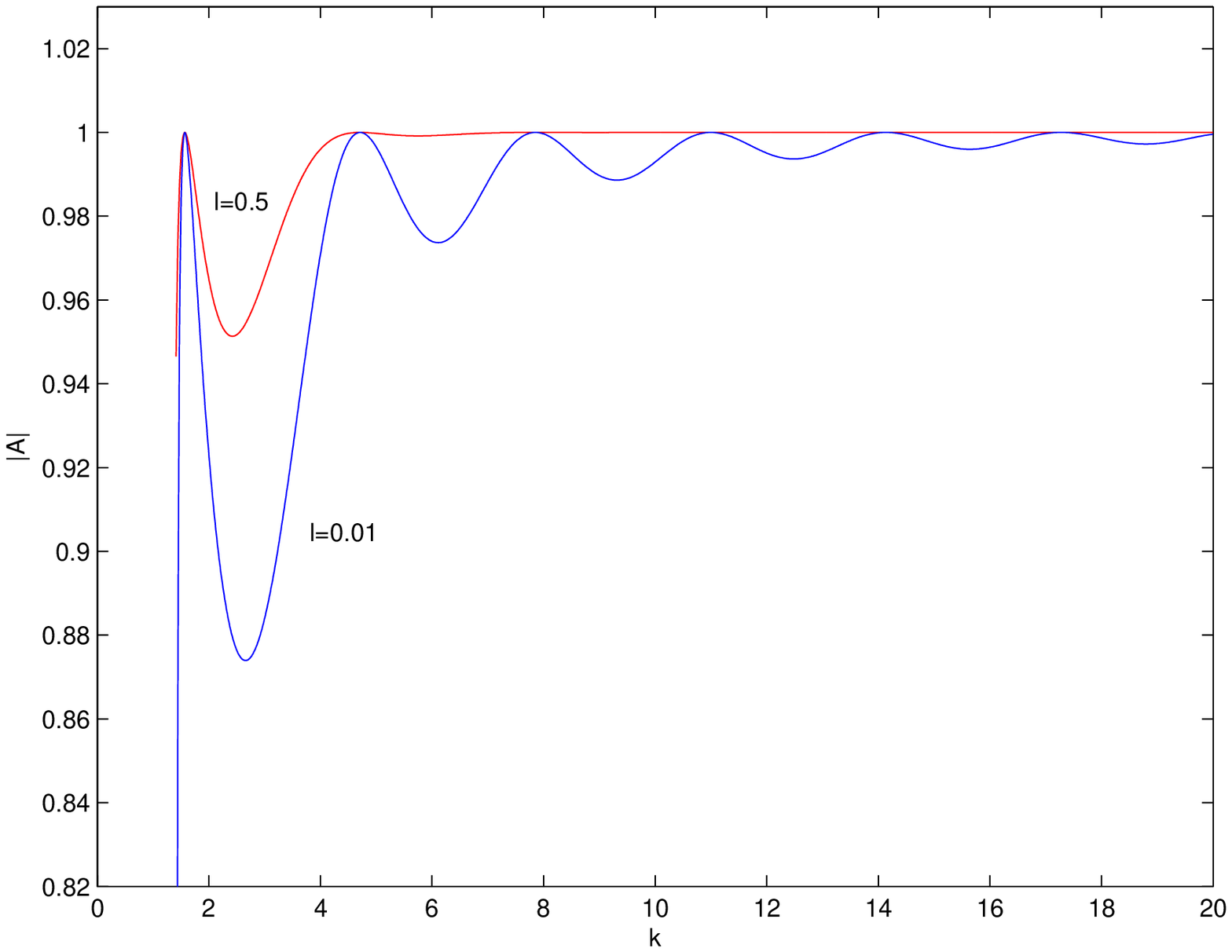,width=11truecm}
\end{center}
\caption{The amplitude $\left\vert
A\right\vert $ for $\ell =0.01$ and $0.5$}
\label{scatampl}
\end{figure}

One sees no evidence for the width of the resonances being quantized in
multiples of $\ell $. Rather, the widths and separation of the resonances is
related to the geometry of the problem. However, what one notices is that as 
$\ell $ approaches the scale of the problem, the resonances become extremely
wide being, in practice, undetectable in the scattering amplitudes. This is
illustrated in Fig.\ref{scatampl} where the amplitude of $A$ is plotted.

Therefore the only effect to be expected is that as soon as one
deals with phenomena close to the scale of $\ell $ only few or no resonances
will be observed. Hadronic resonances being of order $10^{-24}s$, this
establishes an upper limit $\ell \lesssim 10^{-25}s$.

\section{Phases and interference}

In the algebra (\ref{I.3}) a complete momentum space description is possible
with the variables $\left( t,p^{0},p^{1},p^{2},p^{3},\Im \right) $ in the
commuting basis $\left( t,p^{1},p^{2},p^{3}\right) $ with the representation 
$\left( \epsilon =+1\right) $%
\begin{equation}
\begin{array}{lll}
x^{0} & = & t \\ 
p^{0} & = & \frac{i}{\ell }r\sinh \left( \ell \frac{d}{dt}\right) \\ 
\mathcal{\Im } & = & r\cosh \left( \ell \frac{d}{dt}\right)%
\end{array}
\label{PI.1}
\end{equation}

Because of the noncommutativity of time with the space coordinates, the
simpler approach is to consider in each case the eigenvalues of the
Hamiltonian and then to obtain the time evolution of each eigenstate.

From the equation 
\begin{equation}
p^{0}\psi =H\psi   \label{PI.2}
\end{equation}%
one obtains, in first approximation\footnote{%
As seen before (Eq.\ref{sw6}), the time shift generator has corrections of
order $\ell ^{2}$. Therefore this Schr\"{o}dinger-like equation is only an
approximation. However, because in this section one only wants to explore
the effects of the repalcement of derivatives by finite-differences, this
approximation will suffice.}, the time evolution of an eigenstate by 
\begin{equation}
p^{0}\psi _{E}=E\psi _{E}  \label{PI.3}
\end{equation}%
\begin{equation}
\frac{i}{\ell }r\sinh \left( \ell \frac{d}{dt}\right) \psi _{E}=\frac{i}{%
2\ell }r\left( \psi _{E}\left( t+\ell \right) -\psi _{E}\left( t-\ell
\right) \right) =E\psi _{E}  \label{PI.4}
\end{equation}%
that is, the (momentum-space) Schr\"{o}dinger-like equation becomes a
finite-difference equation, as has been conjectured by several authors \cite%
{Caldirola1}. Here, however, it is a direct consequence of the algebra (\ref%
{I.3}). For the case $\epsilon =-1$ the corresponding equation would be 
\begin{equation}
\frac{1}{2\ell }r\left( \psi _{E}\left( t+i\ell \right) -\psi _{E}\left(
t-i\ell \right) \right) =E\psi _{E}  \label{PI.5}
\end{equation}%
The solution of (\ref{PI.4}) is 
\begin{equation}
\psi _{E}\left( t\right) =\exp \left\{ -\frac{it}{\ell }\sin ^{-1}\left( 
\frac{\ell E}{r}\right) \right\} \psi _{E}\left( 0\right)   \label{PI.6}
\end{equation}%
The inverse trigonometric nature of exponentials of this type, have been
interpreted by some authors as meaning that in the quantized time case there
is an upper bound $\frac{1}{\ell }$ for the energy. This is an unwarranted
conclusion because the most general representation of the subalgebra $%
\left\{ p_{0},x_{0},\Im \right\} $ allows for the arbitrary factor $r$.
Therefore the maximum energy (of stationary states) would be $\frac{r}{\ell }
$ and not $\frac{1}{\ell }$. Notice that as soon as one considers also the
space coordinates, $r$ is a variable needed for the consistency of the
representation with the commutation relations (\ref{I.3}) (see the
Appendix). In any case, the fact that there would not be any stationary
eigenstates ($\left\vert \psi _{E}\left( t\right) \right\vert =\left\vert
\psi _{E}\left( 0\right) \right\vert ,\forall t$) for $E>\frac{r}{\ell }$
does not mean that the spectrum of $p^{0}$ has an upper bound.

Consider now the interference pattern of two eigenstates%
\begin{eqnarray}
\left\vert \psi _{E_{1}}\left( t\right) +\psi _{E_{2}}\left( t\right)
\right\vert ^{2} &=&\left\vert \psi _{E_{1}}\left( t\right) \right\vert
^{2}+\left\vert \psi _{E_{2}}\left( t\right) \right\vert ^{2}  \notag \\
&&+2\psi _{E_{1}}\left( 0\right) \psi _{E_{2}}\left( 0\right) \cos \left\{ 
\frac{t}{\ell }\left[ \sin ^{-1}\left( \frac{\ell E_{1}}{r}\right) -\sin
^{-1}\left( \frac{\ell E_{2}}{r}\right) \right] \right\}   \label{PI.8}
\end{eqnarray}%
For a small enough difference $\Delta $ between two large energies, the
oscillation frequency $\omega $ of the interference would be largely
affected by the noncommutative structure. Let%
\begin{eqnarray*}
\frac{E_{1}}{r} &=&\xi +\Delta  \\
\frac{E_{2}}{r} &=&\xi 
\end{eqnarray*}%
Then, the oscillating frequency in (\ref{PI.8}) is%
\begin{eqnarray*}
\omega  &=&\frac{1}{\ell }\left[ \sin ^{-1}\left( \ell \left( \xi +\Delta
\right) \right) -\sin ^{-1}\left( \ell \xi \right) \right]  \\
&\simeq &\Delta +\frac{\ell ^{2}}{6}\left( 3\xi ^{2}\Delta +3\xi \Delta
^{2}+\Delta ^{3}\right) 
\end{eqnarray*}%
which for small $\Delta $ leads to a correction $\Delta \left( 1+\frac{1}{2}%
\ell ^{2}\xi ^{2}\right) $.

As in phase space suppression (or dilation) and particle multiplicity
effects (Sect.2), observation of this correction depends on the product $%
\ell \xi $. For instance $\ell \xi \sim O\left( 1\right) $ for $\ell \sim
10^{-27}s$ and $\xi \sim 300$ GeV.

\section{Electron spin resonance}

Consider an (unpaired) electron interacting with a magnetic field, for which
one considers only its spin degree of freedom . In the basis where $\chi
_{+}\left( t\right) $ and $\chi _{-}\left( t\right) $ are the up and down
spin states the Hamiltonian is%
\begin{equation}
H=\frac{g}{2}\mu _{B}\left( 
\begin{array}{cc}
B_{z} & B_{x}-iB_{y} \\ 
B_{x}+iB_{y} & -B_{z}%
\end{array}%
\right)   \label{ESR.1}
\end{equation}%
Let $B_{z}=B_{0}$ be fixed and $B_{x},B_{y}$ time-dependent. The time
dependence for a massless field $\phi $ is obtained from%
\begin{equation}
\left( \left( p^{0}\right) ^{2}-\left\vert p\right\vert ^{2}\right) \phi =0
\label{ESR.2}
\end{equation}%
Use the commuting basis $\left( t,p^{1},p^{2},p^{3}\right) $ and assume $%
\phi $ to be an eigenstate of momentum%
\begin{equation*}
\left\vert p\right\vert ^{2}\phi =k^{2}\phi 
\end{equation*}%
Then from (\ref{A.9}) and (\ref{A.10}) it follows that Eq.(\ref{ESR.2})
becomes%
\begin{equation}
-\frac{\gamma ^{2}}{\ell ^{2}}\sinh ^{2}\left( \ell \frac{\partial }{%
\partial t}\right) \phi _{+}=k^{2}\phi _{+}  \label{ESR.3a}
\end{equation}%
for $\epsilon =+1$ and%
\begin{equation}
-\frac{\gamma ^{2}}{\ell ^{2}}\sin ^{2}\left( \ell \frac{\partial }{\partial
t}\right) \phi _{-}=k^{2}\phi _{-}  \label{ESR.3b}
\end{equation}%
for $\epsilon =-1$, with solutions (setting $\gamma =1$ which is simply a
momentum unit) 
\begin{equation}
\phi _{+}\left( t\right) =\phi _{+}\left( 0\right) \exp \left( \pm \frac{it}{%
\ell }\sin ^{-1}\left( \ell k\right) \right)   \label{ESR.4a}
\end{equation}%
\begin{equation}
\phi _{-}\left( t\right) =\phi _{-}\left( 0\right) \exp \left( \pm \frac{it}{%
\ell }\sinh ^{-1}\left( \ell k\right) \right)   \label{ESR.4b}
\end{equation}%
the main modification of the noncommutative structure being that $k$ is no
longer the frequency of the massless matter wave,%
\begin{eqnarray*}
\omega _{+} &=&\frac{1}{\ell }\sin ^{-1}\left( \ell k\right)  \\
\omega _{-} &=&\frac{1}{\ell }\sinh ^{-1}\left( \ell k\right) 
\end{eqnarray*}

\begin{figure}[htb]
\begin{center}
\psfig{figure=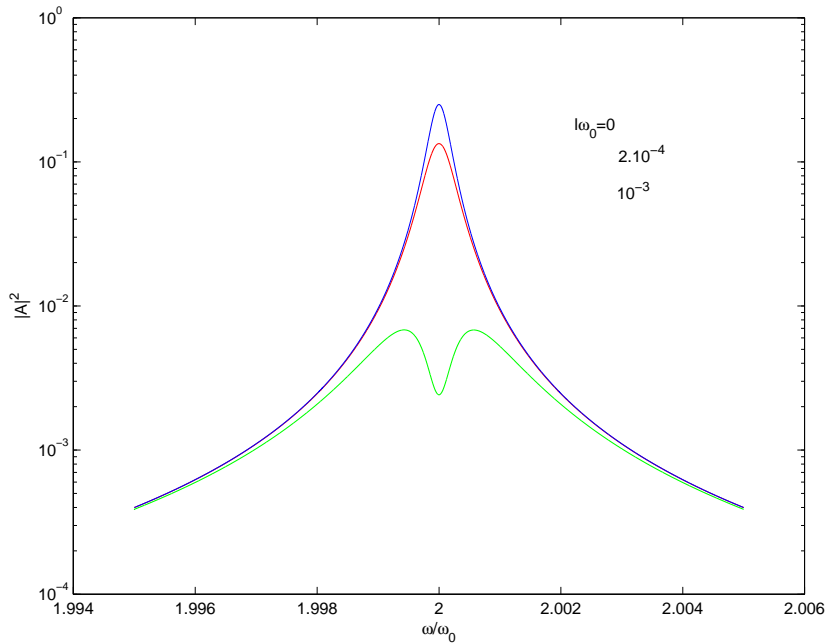,width=11truecm}
\end{center}
\caption{Absorption spectrum for several $%
\ell \omega _{0}$ values}
\label{ESR}
\end{figure}

Consider now a field%
\begin{equation}
B_{x}=b\cos \left( \omega t\right) ;\hspace{0.5cm}B_{y}=b\sin \left( \omega
t\right) ;\hspace{0.5cm}B_{z}=B_{0}  \label{ESR.6}
\end{equation}%
Defining%
\begin{equation}
\omega _{1}=\frac{g}{2}\mu _{B}b;\hspace{0.5cm}\omega _{0}=\frac{g}{2}\mu
_{B}B_{0}  \label{ESR.7}
\end{equation}%
from%
\begin{equation}
p^{0}\chi \left( t\right) =H\chi \left( t\right)   \label{ESR.8}
\end{equation}%
one obtains $\left( \epsilon =+1\right) $%
\begin{eqnarray}
\frac{i}{2\ell }\left\{ \chi _{+}\left( t+\ell \right) -\chi _{+}\left(
t-\ell \right) \right\}  &=&\omega _{0}\chi _{+}\left( t\right) +\omega
_{1}e^{-i\omega _{+}t}\chi _{-}\left( t\right)   \notag \\
\frac{i}{2\ell }\left\{ \chi _{-}\left( t+\ell \right) -\chi _{-}\left(
t-\ell \right) \right\}  &=&-\omega _{0}\chi _{-}\left( t\right) +\omega
_{1}e^{i\omega _{+}t}\chi _{+}\left( t\right)   \label{ESR.9}
\end{eqnarray}%
Replacing $\chi _{-}\left( t\right) $, taken from the first equation, on the
second one obtains with%
\begin{equation}
\chi _{+}\left( t\right) =\chi _{+}\left( 0\right) \exp \left( i\lambda
t\right)   \label{ESR.10}
\end{equation}%
the characteristic equation%
\begin{equation}
\frac{1}{2\ell ^{2}}\left\{ \cos \left( \ell \left( \omega +2\lambda \right)
\right) -\cos \left( \ell \omega \right) \right\} =\frac{\omega _{0}}{\ell }%
\left\{ \sin \left( \ell \left( \omega +\lambda \right) \right) -\sin \left(
\ell \lambda \right) \right\} -\omega _{0}^{2}-\omega _{1}^{2}
\label{ESR.11}
\end{equation}%
which for $\ell =0$ reduces to%
\begin{equation}
\lambda ^{2}+\omega _{+}\lambda +\omega _{0}\omega _{+}-\omega
_{0}^{2}-\omega _{1}^{2}=0  \label{ESR.12}
\end{equation}%
with solution%
\begin{equation}
\lambda _{\pm }^{\left( 0\right) }=-\frac{\omega _{+}}{2}\pm \sqrt{\omega
_{1}^{2}+\left( \omega _{0}-\frac{\omega _{+}}{2}\right) }  \label{ESR.13}
\end{equation}%
To obtain the leading $\ell ^{2}$ corrections to this result one finds from (%
\ref{ESR.11})%
\begin{equation}
\left. \frac{d\lambda }{d\ell }\right\vert _{\ell =0}=0;\hspace{0.5cm}\left. 
\frac{d\lambda }{d\ell ^{2}}\right\vert _{\ell =0}=\frac{\omega _{0}\left(
\left( \omega _{+}+\lambda \right) ^{3}-\lambda ^{3}\right) }{6\left( \omega
_{+}+2\lambda \right) }  \label{ESR.14}
\end{equation}%
Therefore, in order $\ell ^{2}$%
\begin{equation}
\lambda _{\pm }^{\left( 1\right) }=\lambda _{\pm }^{\left( 0\right) }+\ell
^{2}\frac{\omega _{0}\left( \left( \omega _{+}+\lambda _{\pm }^{\left(
0\right) }\right) ^{3}-\lambda _{\pm }^{\left( 0\right) 3}\right) }{6\left(
\omega _{+}+2\lambda _{\pm }^{\left( 0\right) }\right) }  \label{ESR.15}
\end{equation}%
Let the initial conditions at $t=0$ be%
\begin{equation*}
\chi \left( 0\right) =\left( 
\begin{array}{l}
0 \\ 
1%
\end{array}%
\right) 
\end{equation*}%
Then from%
\begin{equation}
\chi _{+}\left( t\right) =Ae^{i\lambda _{+}t}+Be^{i\lambda _{-}t}
\label{ESR.16}
\end{equation}%
$B=-A$ and%
\begin{equation}
\chi _{-}\left( 0\right) =-\frac{A}{\ell \omega _{1}}\left\{ \sin \left(
\ell \lambda _{+}\right) -\sin \left( \ell \lambda _{-}\right) \right\} =1
\label{ESR.17}
\end{equation}%
leading to%
\begin{equation}
\left\vert A\right\vert =\frac{\omega _{1}}{\frac{1}{\ell }\left\{ \sin
\left( \ell \lambda _{+}\right) -\sin \left( \ell \lambda _{-}\right)
\right\} }  \label{ESR.18}
\end{equation}%
If the energy of the electromagnetic field at frequency $\omega _{+}$ is
dissipated by relaxation processes the absorbed energy would be proportional
to $\left\vert A\right\vert ^{2}$. Fig.\ref{ESR} illustrates, for several $%
\ell \omega _{0}$ values, the kind of deviations in the absorption spectrum
that would be observed. Similar results are obtained for the $\epsilon =-1$
case. 

\section{Non-commutative QED}

The general construction of gauge fields as Lie algebra-valued connections
on the deformed algebra (\ref{I.3}) was sketched in Ref.\cite{Vilela3}.
Here, I will review those results and construct the electromagnetic field in
the operator context. Then an operator symbol (star-product) formulation
will be developed which is useful for practical calculations.

\subsection{Non-commutative space-time and the electromagnetic field}

The derivations of the algebra $\mathcal{R}_{\ell ,\infty }$ (\ref{I.3}) are
the inner derivations plus a dilation $\mathcal{D}$ ($[\mathcal{D},P_{\mu
}]=P_{\mu };[\mathcal{D},\Im ]=\Im ;[\mathcal{D},M_{\mu \nu }]=[\mathcal{D}%
,x_{\mu }]=0$).%
\begin{equation}
\text{Der}\{\Re _{\ell ,\infty }\}=\{X_{\mu },M_{\mu \nu },P_{\mu },\Im ,%
\mathcal{D}\}  \label{NQED1}
\end{equation}%
Because in the construction of the differential algebra, the derivations
corresponding to $\frac{1}{i}P_{\mu }$ and $\frac{1}{i\ell }\Im $ play a
special role, they are denoted by the symbols $\partial _{\mu }$ and $%
\partial _{4}$ to emphasize their role as elements of Der$\{\Re _{\ell
,\infty }\}$ rather than elements of the enveloping algebra\footnote{%
For the generation of the enveloping algebra one adds the operators $\left\{
\Im ^{-1},1\right\} $ and their powers. Because $\Im $ is a small
deformation of $1$, $\Im ^{-1}$ is well defined. The commutation relations
with $\Im ^{-n}$ are easily obtained from the vanishing of all commutators
with $\Im \Im ^{-1}$. For example $\left[ X_{\mu },\Im ^{-1}\right] =-i\ell
^{2}P_{\mu }\Im ^{-2}$.} $U_{\Re }=\{X_{\mu },M_{\mu \nu },P_{\mu },\Im ,\Im
^{-1},1\}$ of $\Re _{\ell ,\infty }$. The action on the generators is 
\begin{equation}
\begin{array}{lll}
\partial _{\mu }(X_{\nu }) & = & \eta _{\mu \nu }\Im  \\ 
\partial _{4}(X_{\mu }) & = & \ell P_{\mu } \\ 
\partial _{\sigma }(M_{\mu \nu }) & = & \eta _{\sigma \mu }P_{\nu }-\eta
_{\sigma \nu }P_{\mu } \\ 
\partial _{\mu }(P_{\nu }) & = & \partial _{\mu }(\Im )=\partial _{\mu }(1)=0
\\ 
\partial _{4}(M_{\mu \nu }) & = & \partial _{4}(P_{\mu })=\partial _{4}(\Im
)=\partial _{4}(1)=0%
\end{array}
\label{NQED2}
\end{equation}%
The set of derivations $\{\partial _{\mu },\partial _{4}\}$ is the minimal
set that contains the usual $\partial _{\mu }$'s, is maximal abelian and is
action closed on the coordinate operators, in the sense that the action of $%
\partial _{\mu }$ on $x_{\nu }$ leads to the operator $\Im $ that
corresponds to $\partial _{4}$ and conversely. Denoting by $V$ the complex
vector space of derivations spanned by $\{\partial _{\mu },\partial _{4}\}$,
the algebra of differential forms $\Omega (U_{\Re })$ is now constructed
from the complex $C(V,U_{\Re })$ of multilinear antisymmetric mappings from $%
V$ to $U_{\Re }$. For an explicit construction of $\Omega (U_{\Re })$ one
may use a basis of 1-forms $\{\theta ^{\mu },\theta ^{4}\}$ defined by 
\begin{equation}
\theta ^{a}(\partial _{b})=\delta _{b}^{a},\text{ }a,b\in (0,1,2,3,4)
\label{NQED3}
\end{equation}%
The elements $\theta ^{a}$ of the 1-form basis do not coincide with $dx_{\mu
}$. Actually 
\begin{equation}
dX_{\mu }=\eta _{\nu \mu }\Im \theta ^{\nu }+\ell P_{\mu }\theta ^{4}
\label{NQED4}
\end{equation}%
Although the operators associated to the coordinates are just the four $%
X_{\mu }$, $\mu \in (0,1,2,3)$, (no extra dimension in the set of physical
coordinates) one sees that an additional degree of freedom appears in the
set of derivations which, by duality, leads to an additional degree of
freedom in the exterior algebra. Therefore quantum fields that are
connections may pick up additional components.

To define gauge fields in this setting consider a right $U_{\Re }$-module
generated by $1$. 
\begin{equation}
E=\{1a;\ a\in U_{\Re }\}  \label{NQED5}
\end{equation}%
A connection is a mapping $\nabla :E\rightarrow E\otimes \Omega ^{1}(U_{\Re
})$ such that 
\begin{equation}
\nabla (\chi a)=\chi da+\nabla (\chi )a  \label{NQED6}
\end{equation}%
$\chi \in E$, $a\in U_{\Re }$. For each derivation $\delta _{i}\in V$ the
connection defines a mapping $\nabla _{\delta _{i}}:E\rightarrow E$. Because
of Eq.(\ref{NQED6}), knowing how the connection acts on the algebra unit $1$%
, one has the complete action. Define 
\begin{equation}
\nabla (1)\doteq A=A_{i}\theta ^{i},\ A_{i}\in U_{\Re }  \label{NQED7}
\end{equation}

A gauge transformation will be a unitary element ($U^{\ast }U=1$) acting on $%
E$. Such unitary elements exist in the $C^{\ast }$-algebra formed from the
elements of the enveloping algebra by the standard techniques. Let $\phi \in
E$ be a scalar field. Then 
\begin{equation}
\nabla (\phi )=d\phi +\nabla (1)\phi  \label{NQED8}
\end{equation}%
Acting on $\nabla (\phi )$ with a unitary element 
\begin{equation}
U\nabla (\phi )=Ud(U^{-1}U\phi )+U\nabla (1)U^{-1}U\phi =d(U\phi
)+\{U(dU^{-1})+U\nabla (1)U^{-1}\}U\phi =\nabla ^{^{\prime }}(U\phi )
\label{NQED9}
\end{equation}%
Therefore the gauge field transformation under a gauge transformation is 
\begin{equation}
\nabla (1)\rightarrow U(dU^{-1})+U\nabla (1)U^{-1}  \label{NQED10}
\end{equation}%
The non-commutativity of $U_{\Re }$ prevents the vanishing of the second
term.

The connection is extended to a mapping $E\otimes \Omega (U_{\Re
})\rightarrow E\otimes \Omega (U_{\Re })$ by 
\begin{equation}
\nabla (\phi \alpha )=\nabla (\phi )\alpha +\phi d\alpha  \label{NQED11}
\end{equation}%
$\phi \in E$ and $\alpha \in \Omega (U_{\Re })$. Computing $\nabla ^{2}(1)$%
\begin{equation}
\begin{array}{c}
\nabla ^{2}(1)=\nabla (1A_{i}\theta ^{i})=\nabla (1A_{i})\theta
^{i}+1A_{i}d\theta ^{i} \\ 
=1dA_{i}\theta ^{i}+\nabla (1)A_{i}\theta ^{i}+1A_{i}d\theta ^{i} \\ 
=\partial _{j}(A_{i})\theta ^{j}\wedge \theta ^{i}+A_{j}A_{i}\theta
^{j}\wedge \theta ^{i}%
\end{array}
\label{NQED12}
\end{equation}%
Therefore, given an electromagnetic potential $A=A_{i}\theta ^{i}$ ($%
A_{i}\in U_{\Re }$) the corresponding electromagnetic field is $F_{ij}\theta
^{i}\wedge \theta ^{j}$ where 
\begin{equation}
F_{ij}=\partial _{i}(A_{j})-\partial _{j}(A_{i})+[A_{i},A_{j}]
\label{NQED13}
\end{equation}%
$F_{ij}\in U_{\Re }$. Unlike the situation in commutative space-time, the
commutator term does not vanish and pure electromagnetism is no longer a
free theory, because of the quadratic terms in $F_{ij}$. Also the indices in
the connections (\ref{NQED7}) and gauge fields (\ref{NQED13}) run over $%
(0,1,2,3,4)$, which resulted from the most natural choice for the
differential algebra basis.

To construct an action for the electromagnetic field consider a diagonal
metric $\eta _{ab}=(1,-1,-1,-1,1)$ and construct 
\begin{equation}
G=G_{knl}\theta ^{k}\wedge \theta ^{n}\wedge \theta ^{l}  \label{NQED14}
\end{equation}%
where $G_{knl}=\epsilon _{\cdot \cdot knl}^{ij}F_{ij}\in U_{\Re }$. The
action $S_{A}$ is obtained from the trace of $F\wedge G$%
\begin{equation}
S_{A}=\text{Tr}\{F_{ab}F^{ab}\}=\text{Tr}\{F_{\mu \nu }F^{\mu \nu }+2F_{4\mu
}F^{4\mu }\}  \label{NQED15}
\end{equation}%
$\mu ,\nu \in (0,1,2,3)$.

To discuss matter fields one needs spinors, and an appropriate set of $%
\gamma $ matrices to contract the derivations $\partial _{a}$. A massless
action term for spinor matter fields may be written 
\begin{equation}
S_{\psi }=i\overline{\psi }\gamma ^{a}\partial _{a}\psi  \label{NQED16}
\end{equation}%
where $a\in (0,1,2,3,4)$, $\gamma ^{a}=(\gamma ^{0},\gamma ^{1},\gamma
^{2},\gamma ^{3},i\gamma ^{5})$ and $\psi $ is a field in a projective
module $E_{\psi }\subset $ $U_{\Re }^{\otimes 4}$. It follows from the
properties of the derivations that this term is Lorentz invariant. Notice
that although the set $\{M_{\mu \nu },X_{\mu }\}$ has a O(2,3) structure, it
is only the O(1,3) part that is a symmetry group. Coupling the fermions to
the gauge field 
\begin{equation}
S_{\psi }=\overline{\psi }i\gamma ^{a}(\partial _{a}+igA_{a})\psi
\label{NQED17}
\end{equation}%
One sees that the fermions may be coupled to the connection $A_{a}$ without
having to introduce new degrees of freedom in the fermion sector.

Given the connection $A_{\mu }$ as a member of the enveloping algebra $%
U_{\Re }$ it may be decomposed into a set of operator eigenvalues of the
momenta $\left\{ P_{\mu }\right\} $ with c-number coefficients $A_{\mu
}\left( k\right) $. One has%
\begin{equation}
\left[ P_{\mu },e^{-\frac{i}{2}k_{\nu }\left\{ X^{\nu },\Im ^{-1}\right\}
_{+}}\right] =k_{\mu }e^{-\frac{i}{2}k_{\nu }\left\{ X^{\nu },\Im
^{-1}\right\} _{+}}  \label{NQED18}
\end{equation}%
Then%
\begin{equation}
A_{\mu }=\int d^{4}k\left\{ A_{\mu }\left( k\right) e^{-\frac{i}{2}k_{\nu
}\left\{ X^{\nu },\Im ^{-1}\right\} _{+}}+A_{\mu }^{\dagger }\left( k\right)
e^{\frac{i}{2}k_{\nu }\left\{ X^{\nu },\Im ^{-1}\right\} _{+}}\right\}
\label{NQED19}
\end{equation}%
For the electromagnetic field $F_{\mu \nu }$ in (\ref{NQED13}) one has to
compute%
\begin{equation}
\left[ e^{-\frac{i}{2}k_{\nu }\left\{ X^{\nu },\Im ^{-1}\right\} _{+}},e^{-%
\frac{i}{2}q_{\mu }\left\{ X^{\mu },\Im ^{-1}\right\} _{+}}\right]
\label{NQED20}
\end{equation}%
which in leading $\ell ^{2}-$order is%
\begin{equation}
i\frac{\ell ^{2}}{2}e^{-\frac{i}{2}\left( k_{\nu }+q_{\nu }\right) \left\{
X^{\nu },\Im ^{-1}\right\} _{+}}\left( k_{\nu }q_{\mu }-q_{\nu }k_{\mu
}\right) \Im ^{-2}\Sigma ^{\nu \mu }  \label{NQED21}
\end{equation}%
$\Sigma ^{\nu \mu }$ being a spin operator%
\begin{equation}
\Sigma ^{\nu \mu }=M^{\nu \mu }-\left\{ X^{\nu }\Im ^{-1},P^{\mu }\right\}
_{+}+\left\{ X^{\mu }\Im ^{-1},P^{\nu }\right\} _{+}  \label{NQED21a}
\end{equation}%
Then%
\begin{eqnarray}
F_{\mu \nu }\left( k\right) &=&-i\left( k_{\mu }A_{\nu }\left( k\right)
-k_{\nu }A_{\mu }\left( k\right) \right)  \notag \\
&&+i\ell ^{2}\left\{ \int d^{4}qA_{\mu }\left( k-q\right) A_{\nu }\left(
q\right) \left( k-q\right) _{\sigma }q_{\varepsilon }\Sigma ^{\sigma
\varepsilon }\right.  \notag \\
&&+\left. \int d^{4}qA_{\mu }\left( k+q\right) A_{\nu }^{\dagger }\left(
q\right) \left( k+q\right) _{\sigma }q_{\varepsilon }\Sigma ^{\varepsilon
\sigma }\right\}  \label{NQED22}
\end{eqnarray}%
the last term in (\ref{NQED22}) being the momentum space image of the
noncommuting $[A_{i},A_{j}]$ in (\ref{NQED13}). One sees that noncommutative
pure QED\ is not a free theory, having nontrivial 3- and 4-photon vertices
of order $\ell ^{2}$ which are spin-dependent.

\subsection{An operator symbol formulation}

An algebra of non-commuting operators may be represented in a space of
functions with a modified (star) product. The general context of this
formulation is described in Appendix B. Here a star-product is found which
reproduces the noncommutative features of the space-time algebra. The
non-commuting algebra that is being represented is%
\begin{equation}
\begin{array}{rcl}
\lbrack \overset{\wedge }{P}_{\mu },\overset{\wedge }{P}_{\nu }] & = & 0 \\ 
\lbrack \overset{\wedge }{X}_{\mu },\overset{\wedge }{X}_{\nu }] & = & 
-i\epsilon \ell ^{2}M_{\mu \nu } \\ 
\lbrack \overset{\wedge }{P}_{\mu },\overset{\wedge }{X}_{\nu }] & = & i\eta
_{\mu \nu }\overset{\wedge }{\Im } \\ 
\lbrack \overset{\wedge }{P}_{\mu },\overset{\wedge }{\Im }] & = & 0 \\ 
\lbrack \overset{\wedge }{X}_{\mu },\overset{\wedge }{\Im }] & = & i\epsilon
\ell ^{2}\overset{\wedge }{P}_{\mu }%
\end{array}
\label{OS1}
\end{equation}%
where the hat symbols are meant to emphasize the non-commuting operator
nature of $\left\{ \overset{\wedge }{P_{\mu }},\overset{\wedge }{X_{\mu }},%
\overset{\wedge }{\Im }\right\} $. These are going to be represented by
functions $\left\{ p_{\mu },x_{\mu },\Im \right\} $ with a star product%
\begin{equation}
G\left( p,x,\Im \right) \ast H\left( p,x,\Im \right) =Ge^{\frac{i}{2}\left( 
\overset{\longleftarrow }{\partial _{p}^{\mu }}\eta _{\mu \nu }\overset{%
\longrightarrow }{\Im \partial _{x}^{\nu }}-\overset{\longleftarrow }{%
\partial _{x}^{\mu }}\eta _{\mu \nu }\Im \overset{\longrightarrow }{\partial
_{p}^{\nu }}\right) -\frac{i\epsilon \ell ^{2}}{2}\left( \overset{%
\longleftarrow }{\partial _{x}^{\mu }}M_{\mu \nu }\overset{\longrightarrow }{%
\partial _{x}^{\nu }}+\overset{\longleftarrow }{\partial _{x}^{\mu }}P_{\mu }%
\overset{\longrightarrow }{\partial _{\Im }}-\overset{\longleftarrow }{%
\partial _{\Im }}P_{\mu }\overset{\longrightarrow }{\partial _{x}^{\mu }}%
\right) }H  \label{OS2}
\end{equation}%
From this one obtains for the electromagnetic field%
\begin{equation}
F_{\mu \nu }\left( x\right) =\partial _{\mu }A_{\nu }\left( x\right)
-\partial _{\nu }A_{\mu }\left( x\right) +A_{\mu }\left( x\right) e^{-\frac{%
i\epsilon \ell ^{2}}{2}\overset{\longleftarrow }{\partial _{x}^{\sigma }}%
M_{\sigma \rho }\overset{\longrightarrow }{\partial _{x}^{\rho }}}A_{\nu
}\left( x\right) -A_{\nu }\left( x\right) e^{-\frac{i\epsilon \ell ^{2}}{2}%
\overset{\longleftarrow }{\partial _{x}^{\sigma }}M_{\sigma \rho }\overset{%
\longrightarrow }{\partial _{x}^{\rho }}}A_{\mu }\left( x\right) 
\label{OS3}
\end{equation}%
in $\ell ^{2}-$order%
\begin{equation}
F_{\mu \nu }\left( x\right) =\partial _{\mu }A_{\nu }\left( x\right)
-\partial _{\nu }A_{\mu }\left( x\right) -i\epsilon \ell ^{2}\partial
^{\sigma }A_{\mu }\left( x\right) \partial ^{\rho }A_{\nu }\left( x\right)
M_{\sigma \rho }  \label{OS4}
\end{equation}%
and in momentum space%
\begin{eqnarray}
F_{\mu \nu }\left( k\right)  &=&-i\left( k_{\mu }A_{\nu }\left( k\right)
-k_{\nu }A_{\mu }\left( k\right) \right)   \notag \\
&&+i\epsilon \ell ^{2}\left\{ \int d^{4}qA_{\mu }\left( k-q\right) A_{\nu
}\left( q\right) \left( k-q\right) ^{\sigma }q^{\rho }\right.   \notag \\
&&+\left. \int d^{4}qA_{\mu }\left( k+q\right) A_{\nu }^{\dagger }\left(
q\right) \left( k+q\right) ^{\sigma }q^{\rho }\right\} M_{\sigma \rho }
\label{OS5}
\end{eqnarray}

In the context of the noncommuting phase-space structure defined in (\ref%
{OS1}) the orbital angular momentum would be represented by $\overset{\wedge 
}{X}_{\mu }\overset{\wedge }{P}_{\nu }-\overset{\wedge }{X}_{\nu }\overset{%
\wedge }{P}_{\mu }$. Therefore it makes sense to interpret $M_{\mu \nu }$ as
the spin operator and one obtains the same $F_{\mu \nu }\left( k\right) $
structure as before.

\begin{figure}[htb]
\begin{center}
\psfig{figure=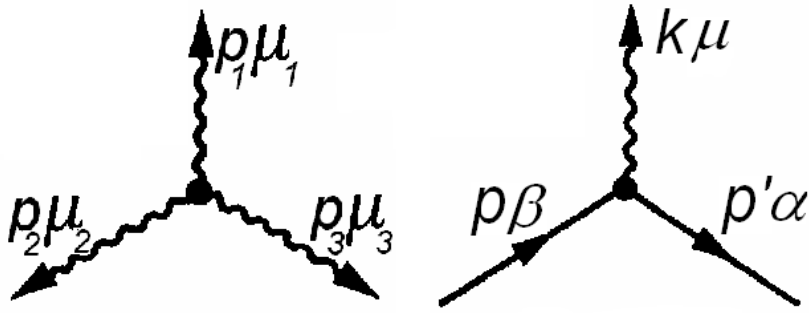,width=11truecm}
\end{center}
\caption{One and three-photon vertices}
\label{vertices}
\end{figure}

For the photon-spinor interactions, by minimal coupling one has%
\begin{equation}
\overset{-}{\psi }\left( x\right) \left( D_{\mu }-m\right) \gamma ^{\mu
}\psi \left( x\right)   \label{OS6}
\end{equation}%
with%
\begin{equation*}
D_{\mu }\psi \left( x\right) =\partial _{\mu }\psi \left( x\right) -iA_{\mu
}\left( x\right) \ast \psi \left( x\right) 
\end{equation*}%
In conclusion: one has 3-photon vertices of order $\ell ^{2}$ and 4-photon
vertices of order $\ell ^{4}$. The 3-photon coupling (in $\ell ^{2}$order) is%
\begin{equation}
\epsilon \ell ^{2}\left( 2\pi \right) ^{4}\delta ^{4}\left(
p_{1}+p_{2}+p_{3}\right) \left\{ g^{\mu _{1}\mu _{2}}p_{1}^{\mu
_{3}}p_{3}^{\sigma }p_{2}^{\rho }M_{\sigma \rho }-g^{\mu _{1}\mu
_{2}}p_{2}^{\mu _{3}}p_{1}^{\sigma }p_{3}^{\rho }M_{\sigma \rho
}+c.p.\right\}   \label{OS7}
\end{equation}%
$c.p.$ meaning cyclic permutations of $\left\{ 1,2,3\right\} $ (refer to Fig.%
\ref{vertices} for notation), the photon-spinor coupling is%
\begin{equation}
e\left( 2\pi \right) ^{4}\delta ^{4}\left( p-p^{^{\prime }}-k\right) \left(
\gamma ^{\mu }\right) _{\alpha \beta }\left\{ 1+\frac{\epsilon \ell ^{2}}{2}%
p^{\prime ^{\sigma }}M_{\sigma \rho }p^{\rho }\right\}   \label{OS8}
\end{equation}%
and the bare propagators are unchanged. All non-commuting contributions have
momentum and spin dependence.

The 3 and 4-photon vertices lead to new one-loop contributions, see Fig.\ref%
{points23} for the 2- and 3-point functions.

\begin{figure}[htb]
\begin{center}
\psfig{figure=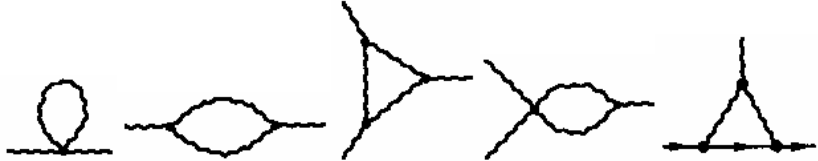,width=11truecm}
\end{center}
\caption{One-loop diagrams}
\label{points23}
\end{figure}

In particular, the new coupling (\ref{OS8})%
\begin{equation*}
-\frac{e\epsilon \ell ^{2}}{2}\overset{-}{\psi }\left( p^{\prime }\right)
k^{\sigma }\gamma ^{\mu }\frac{i}{2}\left[ \gamma _{\sigma },\gamma _{\rho }%
\right] p^{\rho }\psi \left( p\right) 
\end{equation*}%
implies the existence of extra spin-dependent contributions in spinor
scattering which are enhanced at large $p$ and $k$.

\section{Remarks and conclusions}

1) The most relevant point of the stability approach to noncommutative
space-time is the emergence of two deformation parameters, which might
define different length scales. This led to the conjecture that one of them
might be much larger than the Planck length and therefore already detectable
with contemporary experimental means.

2) The effects explored in this paper are rather conservative in the sense
that they explore well-known physical observables. Other similar
consequences of the noncommutative structure, already mentioned before \cite%
{Vilela2}, follow from the non-vanishing right hand side of the double
commutators%
\begin{equation*}
\begin{array}{lll}
\left[ \left[ p_{\mu },x_{\nu }\right] ,x_{\alpha }\right]  & = & \epsilon
\ell ^{2}\eta _{\mu \nu }p_{\alpha } \\ 
\left[ \left[ x_{\mu },x_{\nu }\right] ,x_{\alpha }\right]  & = & \epsilon
\ell ^{2}\left( \eta _{\nu \alpha }x_{\mu }-\eta _{\mu _{\alpha }}x_{\nu
}\right) 
\end{array}%
\end{equation*}

3) A more speculative aspect of the noncommutative structure concerns the
physical relevance of the extra derivation $\partial _{4}$ described in
Section 7.1. This includes new fields associated to gauge interactions which
may lead to effective mass terms for otherwise massless particles (see \cite%
{Vilela3} for more details)

\section{Appendix A: Representations of the deformed algebra and its
subalgebras}

For explicit calculations of the consequences of the non-commutative
space-time algebra (\ref{I.2}) (with $\epsilon ^{\prime }=0$) it is useful
to have at our disposal functional representations of this structure. Such
representations on the space of functions defined on the cone $C^{4}$ ($%
\epsilon =-1$) or $C^{3,1}$ ($\epsilon =+1$) have been described in \cite%
{Vilela3}. Here one collects a few other useful representations of the full
algebra and some subalgebras.

1 - As differential operators in a 5-dimensional commutative manifold $%
M_{5}=\{\xi _{\mu }\}$ with metric $\eta _{aa}=(1,-1,-1,-1,\epsilon )$%
\begin{equation}
\begin{array}{lll}
p_{\mu } & = & i\frac{\partial }{\partial \xi ^{\mu }} \\ 
\Im  & = & 1+i\ell \frac{\partial }{\partial \xi ^{4}} \\ 
M_{\mu \nu } & = & i(\xi _{\mu }\frac{\partial }{\partial \xi ^{\nu }}-\xi
_{\nu }\frac{\partial }{\partial \xi ^{\mu }}) \\ 
x_{\mu } & = & \xi _{\mu }+i\ell (\xi _{\mu }\frac{\partial }{\partial \xi
^{4}}-\epsilon \xi ^{4}\frac{\partial }{\partial \xi ^{\mu }})%
\end{array}
\label{A.7}
\end{equation}

2 - Another global representation is obtained using the commuting set $%
\left( p^{\mu },\Im \right) $, namely%
\begin{equation}
\begin{array}{ccc}
x_{\mu } & = & i\left( \epsilon \ell ^{2}p_{\mu }\frac{\partial }{\partial
\Im }-\Im \frac{\partial }{\partial p^{\mu }}\right)  \\ 
M_{\mu \nu } & = & i\left( p_{\mu }\frac{\partial }{\partial p^{\nu }}%
-p_{\nu }\frac{\partial }{\partial p^{\mu }}\right) 
\end{array}
\label{A.8}
\end{equation}

3 - Representations of subalgebras

Because of non-commutativity only one of the coordinates can be
diagonalized. Here, consider the restriction to one space dimension, namely
the algebra of $\left\{ p^{0},\Im ,p^{1},x^{0},x^{1}\right\} $. 

For $\epsilon =+1$ define hyperbolic coordinates in the plane $\left(
p^{1},\Im \right) $ and polar coordinates in the plane $\left( p^{0},\Im
\right) $. Then, from it follows from (\ref{A.8})%
\begin{equation}
\begin{array}{lll}
p^{1} & = & \frac{r}{\ell }\sinh \mu  \\ 
p^{0} & = & \frac{\gamma }{\ell }\sin \theta  \\ 
\Im  & = & r\cosh \mu =\gamma \cos \theta  \\ 
x^{1} & = & i\ell \frac{\partial }{\partial \mu } \\ 
x^{0} & = & -i\ell \frac{\partial }{\partial \theta }%
\end{array}
\label{A.9}
\end{equation}%
For $\epsilon =-1$ with polar coordinates in the plane $\left( p^{1},\Im
\right) $ and hyperbolic coordinates in the plane $\left( p^{0},\Im \right) $%
,%
\begin{equation}
\begin{array}{lll}
p^{1} & = & \frac{r}{\ell }\sin \theta  \\ 
p^{0} & = & \frac{\gamma }{\ell }\sinh \mu  \\ 
\Im  & = & \gamma \cosh \mu =r\cos \theta  \\ 
x^{1} & = & i\ell \frac{\partial }{\partial \theta } \\ 
x^{0} & = & -i\ell \frac{\partial }{\partial \mu }%
\end{array}
\label{A.10}
\end{equation}

\section{Appendix B: Operator symbol formulation}

Let $\hat{A}$ be an operator in a Hilbert space $\mathcal{H}$ and $\hat{U}(%
\vec{x})$, $\hat{D}(\vec{x})$ two families of operators called \textit{%
dequantizers} and \textit{quantizers}, respectively, such that 
\begin{equation}
\text{Tr}\left\{ \,\hat{U}(\vec{x})\hat{D}(\vec{x}^{\prime })\right\}
=\delta (\vec{x}-\vec{x}^{\prime })  \label{5.2}
\end{equation}%
The labels $\vec{x}$ (with components $x_{1},x_{2},\ldots x_{n}$) are
coordinates in a linear space $V$ where the functions (operator symbols) are
defined. Some of the coordinates may take discrete values. For them the
delta function in (\ref{5.2}) should be understood as a Kronecker delta.
Provided the property (\ref{5.2}) is satisfied, one defines the \textit{%
symbol of the operator} $\hat{A}$ by the formula 
\begin{equation}
f_{A}(\vec{x})=\text{Tr}\left\{ \hat{U}(\vec{x})\hat{A}\right\} ,
\label{5.3}
\end{equation}%
assuming the trace to exist. In view of (\ref{5.2}), one has the
reconstruction formula 
\begin{equation}
\hat{A}=\int f_{A}(x)\hat{D}(\vec{x})\,d\vec{x}  \label{5.4}
\end{equation}%
The role of quantizers and dequantizers may be exchanged. Then 
\begin{equation}
f_{A}^{d}(\vec{x})=\text{Tr}\left\{ \hat{D}(\vec{x})\,\hat{A}\right\}
\label{5.6}
\end{equation}%
is called the dual symbol of $f_{A}(\vec{x})$ and the reconstruction formula
is 
\begin{equation}
\hat{A}=\int f_{A}^{d}(x)\hat{U}(\vec{x})\,d\vec{x}  \label{5.7}
\end{equation}%
Symbols of operators can be multiplied using the star-product kernel as
follows 
\begin{equation}
f_{A}(\vec{x})\star f_{B}(\vec{x})=\int f_{A}(\vec{y})f_{B}(\vec{z})K(\vec{y}%
,\vec{z},\vec{x})\,d\vec{y}\,d\vec{z}  \label{5.9}
\end{equation}%
the kernel being%
\begin{equation}
K(\vec{y},\vec{z},\vec{x})=\text{Tr}\left\{ \hat{D}(\vec{y})\hat{D}(\vec{z})%
\hat{U}(\vec{x})\right\}  \label{5.10}
\end{equation}%
The star-product is associative, 
\begin{equation}
\left( f_{A}(\vec{x})\star f_{B}(\vec{x})\right) \star f_{C}(\vec{x})=f_{A}(%
\vec{x})\star \left( f_{B}(\vec{x})\star f_{C}(\vec{x})\right) ,
\label{5.11}
\end{equation}%
this property corresponding to the associativity of the product of operators
in Hilbert space.

With the dual symbols the trace of an operator may be written in integral
form 
\begin{equation}
\text{Tr}\left\{ \,\hat{A}\hat{B}\right\} =\int f_{A}^{d}(\vec{x})f_{B}(\vec{%
x})\,d\vec{x}=\int f_{B}^{d}(\vec{x})f_{A}(\vec{x})\,d\vec{x}.  \label{5.13}
\end{equation}

For two different symbols $f_{A}(\vec{x})$ and $f_{A}(\vec{y})$
corresponding, respectively, to the pairs ($\hat{U}(\vec{x})$,$\hat{D}(\vec{x%
})$) and ($\hat{U}_{1}(\vec{y})$,$\hat{D}_{1}(\vec{y})$), one has the
relation 
\begin{equation}
f_{A}(\vec{x})=\int f_{A}(\vec{y})K(\vec{x},\vec{y})\,d\vec{y},  \label{5.14}
\end{equation}%
with intertwining kernel 
\begin{equation}
K(\vec{x},\vec{y})=\text{Tr}\left\{ \hat{D}_{1}(\vec{y})\hat{U}(\vec{x}%
)\right\}   \label{5.15}
\end{equation}

This general formulation of operators, as operator symbols in a space of
functions with a star-product, is useful in many other contexts, for example
in signal processing \cite{Briolle3}.

\end{document}